\documentclass[preprint2]{aastex}

\slugcomment{to appear in Astronomical Journal}

\shorttitle{Morphology, Activity and environment of  NGC 3367}

\shortauthors{Hern\'andez-Toledo et al.}

\begin{document}

\title{The Bulgeless Seyfert/LINER Galaxy NGC 3367: Disk, Bar, Lopsidedness and Environment\thanks{Based on data obtained at 
the 0.84m, 1.5m and 2.1m telescopes of the Observatorio Astron\'omico Nacional, San Pedro M\'artir operated 
by the Instituto de Astronom\'{\i}a, Universidad Nacional Aut\'onoma de M\'exico.}}
 
\author{H. M. Hern\'andez-Toledo$^{2,a}$, M. Cano-D\'{\i}az$^{a,b}$, O. Valenzuela$^{a}$, I. Puerari$^{c}$, J. A.
      Garc\'{\i}a-Barreto$^{a}$, E. Moreno-D\'{\i}az$^{a}$, H. Bravo-Alfaro$^{d}$}

      \affil{$^a$Instituto de Astronom\'{\i}a, Universidad Nacional Aut\'onoma de M\'exico, Apartado Postal 70-264,
      Mexico \hbox{D.F.}, 04510, Mexico \\$^b$INAF-Osservatorio Astronomico di Roma, 
     via di Frascati 33, 00040, Monte Porzio Catone, Italy \\$^c$Instituto Nacional de Astrof\'{\i}sica, \'Optica y Electr\'onica, 
     Calle Luis Enrique Erro 1, 72840, Sta. Mar\'{\i}a Tonantzintla, Puebla, Mexico \\$^d$Departamento de Astronom\'{\i}a, 
      Universidad de Guanajuato, Apdo. Postal 144, Guanajuato 36000, Mexico}

\altaffiltext{2}{E-mail: hector@astroscu.unam.mx}

\begin{abstract}

NGC 3367 is a nearby  isolated active galaxy that shows a radio jet, a strong bar and evidence of lopsidedness. We present a quantitative  
analysis of the  stellar and gaseous structure of the galaxy disk and a search for evidence of recent interaction based 
on new  $UBVRI$ H$\alpha$  and  $JHK$ images  and on  archival  H$\alpha$ Fabry-Perot  and HI VLA data. From a coupled 
1D/2D GALFIT bulge/bar/disk decomposition an (B/D $\sim$ 0.07-0.1) exponential pseudobulge is inferred in all the observed 
bands. A $NIR$ estimate of the bar strength $<Q_T^{max}(R)>$ = 0.44 places NGC 3367 
bar among the strongest ones. The asymmetry properties were studied using (1) optical and NIR $CAS$ indexes (2) the stellar (NIR) and gaseous ($H\alpha$, HI) $A_1$ Fourier mode amplitudes and (3) the HI integrated profile and HI mean intensity distribution. While the average stellar component 
shows asymmetry values close to the average found in the Local Universe for isolated galaxies, the young stellar component and gas 
values are largely decoupled showing significantly larger $A_1$ mode amplitudes suggesting that the gas has been recently perturbed. 
NGC 3367 is devoided of HI gas in the central regions where a significant amount of 
molecular CO gas exists instead. Our search for (1) faint stellar structures in the outer regions (up to 
$\mu_R \sim 26 ~mag ~arcsec^{-2}$), (2) ($H\alpha$) star-forming satellite galaxies and (3) regions with 
different colors (stellar populations) along the disk all failed. Such an absence is interpreted using recent numerical 
simulations to constrain a tidal  event with an LMC like galaxy to some dynamical times in the past or  to a current very low mass, gas rich accretion. 
We conclude that a cold accretion mode (gas and small/dark galaxies) may be responsible of the nuclear 
activity and peculiar (young stars and gas) morphology regardless of the highly isolated environment. Black hole growth in bulgeless galaxies may be triggered by cosmic smooth mass accretion.
  
\end{abstract}


\section{Introduction} 

NGC 3367 is a nearby barred galaxy that is classified as SB(rs)c in the Third Reference Catalog of Bright Galaxies (de Vaucouleurs 
et al. 1991, hereafter RC3) and as Sy 2-like, HII in V\'eron-Cetty \& V\'eron (1986). MIR Spitzer observations of NGC 3367 reveal 
the presence of [Ne v] lines at 14.3$\mu$m and 24.3$\mu$m  with X-ray luminosity dominated by a power law with 2-10 keV luminosities of
2.0 $\times 10^{40} erg ~s^{-1}$ (McAlpine et al. 2011).  NGC 3367 is located in the field of the Leo 
Group but it belongs to the background based on its optical velocity of $v_o$ = 2998 ~km s$^{-1}$ and the mean velocity of 
the Leo I Group = 900 ~km ~s$^{-1}$ (Ferguson and Sandage 1990; Stocke et al. 1991; Tonry et al. 2001). The closest candidate for a galaxy 
companion is NGC 3391 at a projected distance of $\sim$ 563 kpc or 18 optical diameters away (Garc\'{\i}a-Barreto et al. 2003) and a recent 
search for isolated galaxies in the local Universe based on the SDSS has also confirmed that NGC 3367 is an isolated galaxy 
(Hern\'andez-Toledo et al. 2010). The optical appearance of NGC 3367 is dominated by a bright bar and an apparent large-scale 
asymmetry or lopsidedness to the southwest side.

Radio observations in NGC 3367 reveal a bipolar synchrotron outflow from the nucleus and two large lobes with a total projected extent on 
the sky (from NE-to-SW) of 12 kpc, resembling a radio galaxy. The axis of the ejected outflow is highly inclined with respect to the axis 
of rotation of the disk (Garc\'{\i}a-Barreto et al. 1998; Garc\'{\i}a-Barreto, Franco \& Rudnick 2002). Single-dish HI content of 
$M_{HI} \sim 7 \times 10^{9} M_{\odot}$ has been reported for this galaxy (Huchtmeier \& Richter 1989) and a high fraction of molecular 
gas has been found mostly concentrated in the central $27''$ (r=5.7 kpc) $M(H2) = 2.7 \times 10^{9} M{\sun}$ (Garc\'{\i}a-Barreto et al. 2005). 
NGC 3367 also shows weak thermal radio continuum emission (at 4.5'' angular resolution) extended throughout the disk (Garc\'{\i}a-Barreto et 
al. 1998) and H$\alpha$ Fabry-Perot observations indicate that its rotation axis lies projected on the disk at a P.A. of $141^{\circ}$, 
the N side of that projected line being closer to the observer (Garc\'{\i}a-Barreto \& Rosado 2001).

Despite that several properties of NGC 3367 somehow resemble a gravitational interacting system with another galaxy, there is yet 
no supporting evidence. This is the first paper of a series where we present our multi-wavelength data and discuss some alternatives to 
the origin of the observed morphology and asymmetries. Our paper is split into two general parts; one describing the observations and 
the estimate of the galaxy parameters and the second discussing the interpretation of measurements as constraints to the recent 
dynamical history in NGC 3367. Specifically, a description of observations and the techniques used in the data reduction are 
briefly described in Section 2. In Section 3 we present a brief summary of the main morphological features found in the optical and NIR 
images. A surface photometry analysis is carried out in Section 4, including (i) an analysis of the surface brightness and color 
profiles, (ii) a coupled 1D-2D bulge/disk and bulge/bar/disk decomposition into S\'ersic, Exponential and Ferrer components. 
Section 5 presents different bar properties like ISM shocks based on the dust lane geometry revealed by a $B-I$ color map, a 
quantification of the bar strength and length, and a brief discussion of the bar nature in terms of the Athanassoula \& Misiriotis 
(2002) models. In Section 6 Lopsidedness is reviewed from various estimators, namely, the Concentration-Asymmetry-Clumpiness 
(hereafter $CAS$) parameters in the optical and NIR bands, the $m = 0-2$ Fourier NIR and H$\alpha$ amplitudes and the VLA HI asymmetry 
from the moment 0 distribution and the corresponding HI line profile. In Section 7 we look for (i) evidence of low surface brightness 
features likely related to a tidal origin and (ii) the presence of recent galaxy accretion events or satellite companions in 
the observed colors. Section 8 presents a general discussion on the origin of the disk Lopsidedness. We use Fabry-Perot 
H$_\alpha$ and HI VLA data to discuss possible evidence of anisotropic gas accretion and its possible connection to the AGN 
activity in NGC 3367. Finally, our summary and concluding remarks are presented in Section 9. A distance of 43.6 Mpc 
($H_{o} = 75 ~km ~s^{-1} Mpc^{-1}$) is adopted for NGC 3367 (Tully 1988), resulting in a linear scale of 210 ~pc ~arcsec$^{-1}$.

\section{Observations and Data Reduction} \label{DataReduction}
\label{sec:obs}
 
The optical $UBVRI$ observations were carried out at the 0.84m telescope of the Observatorio Astron\'omico Nacional at San Pedro 
M\'artir (OAN-SPM), Baja California, M\'exico, with a Site1 CCD detector yielding a total field of view of 7.2 x 7.2 arcmin and 
typical seeing FWHM values of 1.8 arcsec. A detailed description of the optical observations and the standard data reduction within 
the IRAF platform \footnote {IRAF is distributed by the National Optical Astronomy Observatories, which are operated by the 
Association of Universities for Research in Astronomy, Inc., under cooperative agreement with the National Science Foundation} 
can be found in Garc\'{\i}a-Barreto et al. (2007). The routines under Space Telescope Science Data Analysis 
System (STSDAS) were used in the reduction and analysis of both the optical and near-IR data. In this paper a further processing 
of the images in 2 $\times$ 2 binning mode was applied to have a pixel scale of 0.85 arcsec/pix, similar to that in the 
Near-IR observations.

The Near-IR observations were carried out using the CAMILA instrument (Cruz-Gonz\'alez et al. 1994) at the OAN-SPM 2.1m telescope. 
The $CAMILA$ instrument uses a NICMOS 3 detector of 256 x 256 pixel format. The instrument was used in the imaging mode with the focal 
reducer configuration f/4.5 in all our observations, resulting in a spatial scale of 0.85 ~arcsec/pix and a total field of view 
of 3.6 x 3.6 arcmin. The imaging observations were carried out using the broadband $J$, $H$ and $K'$ filters. Each observation 
consisted of a sequence of object and sky exposures, with the integration time of an individual exposure limited by the sky counts, 
which was kept well below the nonlinear regime of the detector. The final exposure times on NGC 3367 were 23, 11 and 21 minutes for 
the $J$, $H$ and $K'$ bands respectively. The photometric calibration of the $JHK$ system was performed using the U.K. Infrared 
System (Hunt et al. 1998). The sky conditions were almost photometric with a typical seeing FWHM of 2 arcsec.

We also dispose of VLA HI data coming from the NRAO VLA-archive, originally obtained in 2001 with the $C$-configuration. We produced
a full Natural Weighting data-cube in order to improve the sensitivity. Our final data-cube has 63 channels and a beam size of 
18.7 $\times$ 15.6 arcsec. We fully describe the HI data reduction in a forthcoming paper (Bravo-Alfaro et al. 2011, in prep).

\section{General Morphology} \label{Morphology} 

The morphology of the galaxy in the optical and NIR bands is shown in Figure 1. Filter-enhanced versions in each band are also 
presented (right-hand side), where the sky has been subtracted and the images have been Gaussian-filtered and 
then subtracted from the original image to enhance both internal and external structures in the form of star forming regions, structures 
embedded into dusty regions or faint outer details of particular interest for our discussion.   

\placefigure{fig-1}

The bluer bands highlight a sharp semi-circular outline at the N-W-S direction. This structure presents a complex pattern of arms 
at a radius of about 50 arcsec (10 kpc) from the center. The arms in the inner region almost coalesce to 
form an internal ring. Some bright HII regions are strung along the inner parts of two major arms that begin at the ends of the bar, 
these arms experiencing various levels of winding and branching.

The contrast produced by the filter-enhanced images enable us to visualize (i) fainter structure beyond 50 arcsec (10 kpc) in the 
form of fragmentary structures at the north and along the external border of the outline (U-to-R band images) but also offers hints 
on the more circular nature of the external disk ($R$ and $I$ bands) and (ii) emphasize a 
falloff of the surface brightness at about 50 arcsec ($U$, $B$ and $V$ bands) that is observed to occur not with the 
same abruptness in the western than in the northeastern side.

The NIR band images show the prominence of the bar as it is clearly enhanced by a dominance of an old population of 
stars in that region. There are also some localized high surface brightness knots in the eastern end of the bar, perhaps 
evidencing the presence, depending on their age, of either a non-negligible population of asymptotic giant branch stars 
or red giant and supergiant branch stars there. Notice, as shown in Figure 2 below, 
that the typical depth of the NIR images is $\sim 21 ~mag \,arcsec^{-2}$ while that of the optical images is $\sim 23-24 
~mag \,arcsec^{-2}$ at 60 arcsec from the center, respectively. This is a relevant issue, specially when discussing properties 
of the external disk.

\section{Surface Photometry Analysis} 
\subsection{Surface Brightness and Color Profiles}
\label{sec:profiles}

Azimuthally averaged radial surface brightness (SB) profiles have been obtained from our optical $UBVRI$ images (see Garc\'{\i}a-Barreto 
et al. (2007)) and from our NIR images as an initial characterization of NGC 3367 structure. 
Ellipse fitting to the isophotes was used to estimate radial profiles of intensity, ellipticity ($\epsilon$), and position angle ($P.A.$). 
The ellipses were calculated using the routines under the STSDAS package. Prior to profile extraction, the FWHM of stars in all the images 
were matched to the one having the poorest seeing (2.3 arcsec). The SB profiles were extracted using ellipse fitting with a fixed center. 
In order to assure a homogeneous computation of structural parameters and color gradients, we use the $R$ band external isophotes as input 
parameters to determine the SB profiles in $UBVIJHK$ bands. We trace SB profiles up to $\sim 23-25 mag ~arcsec^{-2}$ in optical 
bands and $\sim 21-22 ~mag ~arcsec^{-2}$ in the NIR bands. These levels correspond to conservative SB errors of $\sim 0.12 ~mag ~arcsec^{-2}$.

The results of the surface photometric analysis are plotted in Figure 2. To avoid crowding, the surface brightness and color profiles 
for a few wavelengths are plotted. Two dashed vertical lines in each panel indicate, from left to 
right, the end of the bar and the position of the NE surface brightness fall-off respectively. The middle panels show the corresponding 
ellipticity and P.A. radial profiles aimed to illustrate the difficulties behind a 
correct identification of the bar properties as a function of wavelength also in the presence of prominent knots at both ends of the bar 
(see Figure 1). The lower panel shows the external ellipticity and P.A.radial profiles of NGC 3367 estimated from an additional deep $R$ 
band image (see Figure 11 below) that was built from the sum of our various $R$ band exposures adding up to a total of 1.4 hours exposure, 
useful to estimate a representative geometry for the external disk.

\placefigure{fig-2}

The upper left panel of Figure 2 shows a systematic flattening of the surface brightness profiles as we go from $K$ to $U$ bands. 
This behavior could be interpreted as a signature of an optically thick disk at the central regions becoming mildly transparent at large 
radii (Evans 1994). The flattening in the SB profile is correlated with a color gradient showing that the disk becomes gradually bluer at 
outer radii. A red bump in the color profiles reaches as red as 4.5 ~mag in ($B - K$) in the first few arcsec and decreases to about 4 ~mag 
at about 17 arcsec which corresponds to the end of the stellar bar (first vertical dashed line). The peak of the red bump is consistent 
with the maximum concentration of CO emission (Garc\'{\i}a-Barreto et al. 2005). In the external region of the surface brightness profile we notice a 
turnover or possible disk truncation at $\sim 55$ arcsec (second vertical dashed line), this truncation being more evident in the optical 
than in the NIR bands. van der Kruit (1988) proposed that many stellar disks may posses radial cutoffs at approximately the same location 
as seen in the star formation regions. Figure 1 shows that this turnover is visible at approximately the same position corresponding to a 
radius where the abrupt falloff in star formation is appreciated.

The bar 
ellipticity was measured as the maximum of the ellipticity within the bar region while the bar length was measured at the transition 
region between the bar and the disk where both the ellipticity and P.A. change significantly. A set of high surface brightness knots 
precisely at the NE and SW ends of the bar were identified as the dominant sources influencing a correct estimate of the bar ellipticity 
and P.A. Similarly in the external regions, another set of high surface brightness knots along the N-W-S outline were identified as 
the dominant features influencing a correct estimate of the geometry of the disk. Instead of letting the ellipse-fitting routines run blindly,
we carefully masked all the prominent knots and implemented an interactive routine to superpose each isophote onto the corresponding images, 
allowing us for a visual checking of the masking performance and the correct estimate of the bar parameters.  

The bar length in the $K$ band converged to $r \sim 17$ arcsec, corresponding to a diameter of $\sim 7$ kpc at the distance 
of NGC 3367. Similarly, the observed P.A. of the bar is $66^{\circ} \pm 4$, both quantities in agreement with other estimates 
(Garc\'{\i}a-Barreto et al. 1996;1998). A representative ellipticity of the bar (corrected for inclination) in the $J$ and $K$ bands 
is $\epsilon_{max} = 0.58$ $\pm 0.05$. Finally, from the lower panel in Figure 2,  we adopt a nominal radius of 80 arcsec ($\sim$ 3.2 $R$ band 
disk scale-lengths) to estimate an inclination of $25^{\circ} \pm 5^{\circ}$ for NGC 3367. 
This value is within the reported range of values coming from different methods, as discussed in Garc\'{\i}a-Barreto \& Rosado (2001) and more 
recently, from a kinematic analysis using various gas tracers (Cano-D\'{\i}az et al. in preparation).

\subsection{1D-2D Bulge/Bar/Disk Decomposition}
 
Although NGC 3367 has a prominent stellar bar, a common approximation is to perform a 
bulge/disk decomposition to an azimuthally averaged one-dimensional profile (in fixed P.A. and  $\epsilon$ mode) that smooths 
out the bar as well as the spiral arms in the profile extraction. In a first-order approach, we fit the radial intensity 
profiles with a composite profile containing a S\'ersic bulge and an exponential disk. Our algorithm splits 1D galaxy luminosity 
profiles into bulge and disk components simultaneously by using a nonlinear Levenberg-Marquardt least-squares fit (Press et al. 1992)  
to the logarithmic intensities. Seeing effects are accounted for by convolving the theoretical bulge-disk SB profiles with a radially 
symmetric Gaussian PSF.

In the upper-left panel of Figure 2, notice that the shape of the $U$ band SB profile resemble a truncated Freeman type II 
profile. However, that shape practically disappears at longer wavelengths in the NIR bands, this fact combined with the high central
CO emission reported by (Garc\'{\i}a-Barreto et al. 2005) suggest that extinction may play a significant role in the gradual change of 
surface brightness profile across wavelengths. We did not take into account that apparent truncation in the $U$ and $B$ bands. Instead, we 
homogenized the 1D fits by using the same radial zones (up to 55 arcsec) in all the $U$ to $K$ band profiles and then the fitting was 
carried out. At this stage, our results indicate that the bulge is exponential in all the observed bands (S\'ersic n = 0.9-1.1).

Next, our images were prepared to run a set of 2D fitting algorithms (GALFIT; Peng et al. 2002; 2010). To those purposes,the fitting 
region, sky background, PSF image and pixel noise map were all estimated and generated. In a first stage, we coupled our 1D method to 
GALFIT by feeding the results of our 1D fits and the STSDAS isophotal analysis ($n$ S\'ersic index, axial ratio, position angle and 
boxiness/diskyness parameters; see Figure 2) as priors to model a 2D generalized S\'ersic bulge and an exponential disk (in free 
fitting parameter mode) just similar to our 1D case. Figure 3 shows our best 1-D fit to the $J$ band surface brightness distribution 
(left panel) as well as the residual image of NGC 3367 after subtracting the best 2-D bulge/disk fit (right panel). The results for the 
disk and the exponential bulge (S\'ersic n = 0.9-1.2) are consistent with our 1-D fits. However, relatively poor $\chi^{2}$ values were 
obtained in all cases. 

In a second stage we carried out a higher order analysis including a Ferrer's bar in addition to the bulge and disk components.
The Ferrer's bar (Binney and Tremaine 1987) has a nearly flat core and an outer truncation. The sharpness of the truncation is governed 
by a parameter $\alpha$, whereas the central slope is controlled by a parameter $\beta$. This profile is defined up to a an outer radius 
$r_{out}$ beyond which the profile has a value of 0. Our procedure assumed as priors both the representative 1D isophotal parameters 
within the bar region (see Figure 2) as well as the parameters obtained in the bulge/disk 2D fitting. Both a simultaneous fitting 
of the bulge and the disk followed by the inclusion of the bar as well as a simultaneous fit to the disk and the bar, followed by 
the bulge fitting were considered. The best values of $\alpha = 0.5-0.7$ and $\beta = 1.89-1.96$ of the Ferrer's bar for a fixed 
truncation radius ($r_{out}$ = 16 arcsec) yielded disk parameters consistent with the ones found in our previous 1D-2D bulge/disk 
decomposition but with a bulge $n$ S\'ersic index between 1 and 1.8 and $r_e \sim$ 2.5-5 arcsec. Figure 3 (lower-right panel) shows 
the residual image of NGC 3367 after subtracting the best 2-D bulge/bar/disk fit ($\chi^{2} \sim 1$). The residuals show almost no bar 
resemblance and still an intricate nuclear structure. A possible explanation is that the bulge have non-regular structure, perhaps an 
unresolved circumnuclear structure as suggested by the (B-V) pixel map in Figure 12, in addition to a central point source (AGN) that 
was not included in the GALFIT models. Notice that the bar residuals may suggest a bar horizontal bending mode, not included in 
our analysis. Another possible source for the residuals could be related to the dust distribution in the inner regions of NGC 3367.
Notice also that the spiral arms are not heavily important inside one disk scale length and thus were not modelled. The range of 
$n$ S\'ersic indexes found for the bulge suggest B/D ratios ranging form 0.07 to 0.11.
  
Our results assess the likelihood of a pseudo-bulge in NGC 3367 consistent with an exponential luminosity distribution in all the observed 
wavelengths, contrary to a classical bulge structure as suggested by Dong et al. (2006). The presence of a strong bar in this galaxy 
makes that a pure bulge/disk fitting procedure be greatly influenced by the radial range over which the fitting is done causing, depending 
on the adopted parameters, possible overestimates of the bulge. It is thus important to consider the bar to characterize the bulge of 
NGC 3367.  
  
\placefigure{fig-3}

\section{Bar Dissection} 

The evolution of galaxies in later stages of the Universe may be governed  by slow secular processes, related to collective 
dynamical phenomena, such as bars, spiral arms  or the dark matter halo response to baryons (Kormendy \& Kennicutt 2004). 
Do bars contain information about the strength of secular evolution? Bars may drive spiral density waves (Kormendy  \& Norman 1979), 
generate resonance rings of stars and gas (Buta \& Combes 1996), change abundance gradients (Martin \& Roy 1994), or induce gas 
inflow that may lead to bar weakening and bulge growth (Norman et al. 1996).

\placefigure{fig-4}

\subsection{Dustlanes} 

The structural differences between $B$ and $I$ band images can be used to map the star formation and dust distributions. 
Figure 4 shows a close-up of the inner region of NGC 3367 as seen through a $B-I$ color map. Dark regions represent redder colors. 
A dust lane, conspicuous in this 
color map, can be appreciated as a dark curved feature running from east to west of the nucleus. The nucleus and part of the 
bar/arms can also be appreciated as gross bluer features. The dust lane appear slightly offset from the bar major axis toward 
its leading edge. Dust lanes are  caused by building up of gas at shocks  (Prendergast 1966; van Albada \& Sanders 1982; 
Prendergast 1983;  Athanassoula 1992). Dust lanes are closely linked to the bar mass, 
potential ellipticity and pattern speed.  Athanassoula (1992) has modeled systematically the  ISM  response in disks to stellar bars 
scanning different masses, structure and angular speed,  concluding that shocks strength and location are useful to infer the bar physical 
properties (Weiner et al. 2001). The dust lane  curvature observed in Figure 4 when compared with the results of Athanassoula (1992), 
suggests that the bar in NGC 3367 might  be either weak or not  as fast as the ones found in early type galaxies which are lines parallel 
to the bar major axis. As a reference, Salo et al. (1999)  presented one of the largest collection of $\Omega_p$ estimates for 38  OSUBSGS  
spiral galaxies  (Rautiainen, Salo \& Laurikainen 2008), concluding that the pattern speed of the bar depends roughly on the morphological
type. The average value of corotation resonance radius to bar radius increases from 1.15 $\pm$ 0.25 in types 
SB0/a–SBab to 1.44 $\pm$ 0.29 in SBb and 1.82 $\pm$ 0.63 in SBbc–SBc types, similar in type to NGC 3367. While the curved dust 
lane morphology in Figure 4 might be consistent with NGC 3367 having a relatively slow bar, it is known that this
diagnostic is degenerated and a weak bar can trigger curved dustlanes (Athanassoula 1992). On the other hand, Gabbasov et 
al. (2009) recently applied the Tremaine \& Weinberg method (Tremaine \& Weinberg 1984) to a set of Fabry-Perot 
H$\alpha$ gas observations in NGC 3367 finding a value of $\Omega_p = 43 \pm$ 6 km $s^{-1} kpc^{-1}$, consistent with a fast bar ($R \leq 1.4$). 
Although currently we do not have enough data to sort out this apparent contradiction, a forthcoming paper (Cano-D\'{\i}az et al. in 
preparation) will treat this issue.

\subsection{Bar Shape and Nature} 

Figure 5 shows intensity profiles along the major and minor axis of the bar. Athanassoula \& Misiriotis (2002)  have suggested that 
the bar structure is sensitive to which is the dominant galaxy component exchanging angular momentum with the bar itself. They 
introduce the following nomenclature for profiles along the bar major axis:  Halo ($MH$)/Disk ($MD$)/Disk-Bulge ($MDB$) in their 
figure 5. The characters in parenthesis indicate the dominant structure in the bar region, namely a centrally concentrated halo (MH), 
a less centrally concentrated halo or disk (MD) or a bulge and a non-centrally concentrated halo or bulged disk (MDB). For instance, 
that indicates the structural component dominating the angular momentum extraction out from the bar.  
Since NGC 3367 is a low inclination galaxy ($\sim 25^{\circ}$, see Section 4.1), projection effects are not considered as an important 
systematic to assign a morphological bar type from the intensity profiles. However, it is important to mention that a direct comparison 
of theoretical predictions with NGC 3367 is subject to knowing stellar M/L gradients and also any dust effect. We minimize the latest by 
using K-band images and adopting a ratio of M/L $\sim$ 1, consistent with the values reported by Faber \& Gallagher (1979) for late-type 
spiral galaxies (see also Bershady et al. 2010)  
\placefigure{fig-5}.

In particular $MH$ models show a flat surface brightness profile along the bar major axis. Evaluating that situation for NGC 3367 
is more complex because of the multiple spiral arms. In Figure 5, 
we can see that there is a lack of a flat feature in the profile making the $MH$ model unlikely. In order to 
decide between $MD$ and $MDB$ models, we must also compare figure 3 in Athanassoula \& Misiriotis (2002) with images in our Figure 1. 
Although the bar in NGC 3367 is strong (see section below), it is by no means similar to the one in $MH$ models, however, it is also 
not as round as the bar in a $MD$ model. We tentatively conclude that NGC 3367 bar is closer to $MDB$ models, but because of the the 
negligible bulge inferred from our B/D analysis, Figure 5 could be evidencing for a subdominant dark matter mass 
inside an exponential length. Notice that in this interpretation the halo contribution is not negligible to the one from the disk, 
this is not un-reasonable because Halo Adiabatic Contraction models predict a similar structure (Klypin et al. 2002; Gnedin et al. 2004; 
Colin et al.  2006). Although it is possible to argue that dust extinction is masking a light profile truly closer to an  $MD$ or $MH$ model, 
notice however that not only the $K$ band but even the 3.6$\mu$ images from Spitzer (Scoville et al. 2008) support our  
conclusion that $MDB/MH$ model represents better the surface density profile. A stronger assessment of the result requires re-examination 
including stellar population and dust radiative transfer analysis and also considering bar properties in an accreting galaxy.

\subsection{Bar Strength} 

The evolution caused by bars in disk galaxies is due to gravity torques. Any disk asymmetry like a bar or spiral mode give rise to 
tangential forces in addition to radial forces, and then to a gravity torques. The computation of the gravitational torques provides 
important information about the average strength of the perturbing potential. Combes \& Sanders (1981) suggested that these torques 
could provide a useful way to quantify  nonaxi-symmetric features strength if the potential could be estimated. The potential can be 
estimated from de-projected $NIR$ images by solving Poisson equations, for example using Fourier transform techniques, along with 
assumptions concerning mass-to-light ratios and vertical density distributions (Quillen et al. 1994). From this 
potential, the radial and tangential components of the forces in the plane of the galaxy can be estimated. Following this ideas 
we estimate the bar strength in NGC 3367 by using the following equation (Combes \& Sanders 1981, Buta \& Block 2001):

\begin{eqnarray}
Q_T(R)={{F_T^{max}(R)}\over{<F_R(R)>}}
\end{eqnarray}

where $F_T^{max}(R)=1/R[\partial \Phi(R, \theta)/\partial \theta]_{max}$ represents the maximum amplitude of the tangential force at 
radius R and $<F_R(R)>=(d\Phi_0/dR)$ is the mean axi-symmetric radial force at the same radius, derived from the $m=0$ component of 
the gravitational potential. The potential $\Phi$ is estimated by using the NIR images, assuming a constant $(M/L)$ ratio across the 
disk. Then the potential in two dimensions can be derived as the convolution of the mass density with the function 1/R using fast Fourier 
transform techniques (Binney \& Tremaine 1987, p. 90). The gravitational torque is a function of R, but the maximum value of $Q_T(R)$ can 
provide a single measure of bar strength for a whole galaxy, if the gravitational potential is known. Table 1 shows the estimates 
of $Q_T(R)$ for NGC 3367 from the $J$, $H$ and $K$ band images. In our analysis we adopt $h_z$ = 325 pc, which is the exponential scale 
height of our Galaxy (Gilmore \& Reid 1983).  The study by de Grijs (1998) indicates that late-type galaxies on average have a thinner 
disk than earlier type systems. To account for possible variations in scale height and for the possibility that some bars are thicker 
than their disks, we have made separate potential runs for $h_z$ = 225 and 425 pc. Figure 6 shows the ratio map for NGC 3367 ($h_z$ = 
325 pc) with the contours from an $R$ band image overploted.

\placetable{tbl-2}
\placefigure{fig-6}

Figure 6 shows the characteristic pattern of a bar with four well-defined regions where the ratio, depending on quadrant relative to 
the bar, can be negative or positive due to the change of sign of the tangential force. The color coding is useful to visualize the 
differences that exist in $Q$ by quadrant. These differences may be explained either by the presence of the inner spiral arms enclosing 
the bar (see Figure 1) or by some intrinsic asymmetry in the bar itself that could make these regions unequal. Table 1 in Buta \& Block 
(2001) defines the bar strength classes that we adopt in our study. NGC 3367 approximately retains its bar strength class independently 
of scale height variations from 225 to 425 pc. We find that an uncertainty of $\pm$ 100 pc in $h_z$ produces an average uncertainty of 
$\sim 5-10\%$ in bar strength. We report that the maximum tangential force reaches 44\% of the mean radial force with a mean value 
of $<Q_T^{max}(R)>$ = 0.44 for $h_z$ = 325 pc. This value places NGC 3367 in the borderline between classes 4 and 5 of Buta \& Block 
bar-strength classification scheme. In that study, classes 1, 2 and 3 indicate a weak bar, while only a small 
fraction of the galaxies have a bar class $=$ 4 or larger suggesting that NGC 3367 can be considered as a galaxy with a strong bar. 
This strong gravitational torque could be the source of secular evolutive processes. The significant increase in molecular gas mass 
toward the central regions of NGC 3367 as reported by Garc\'{\i}a-Barreto et al. (2005) could be related to such processes. Secular 
scenarios invoking bar formation or dissolution and bulge formation (possibly related to the presence of nuclear activity) might be viable 
mechanisms for the evolution of isolated galaxies along the Hubble sequence (Bournaud \& Combes 2002).

Alternatively, bars can be characterized as strong when they have large ellipticities or the tangential forces are large. As the orbital 
families of bars strongly depend on the underlying gravitational potential, a correlation between the ellipticity and the tangential 
force is expected (Laurikainen et al. 2002; Buta et al. 2004). We have compared our de-projected bar ellipticity in NGC 3367 with 
that predicted from the $<Q_T^{max}(R)>$ - $\epsilon$ relation in Laurikainen et al. (2002). A consistent bar strength estimate
could be obtained. The fact that we can directly measure the bar strength and also the pattern speed opens the possibility of accurately 
estimating the central stellar M/L. We are exploring that possibility in a forthcoming paper (Cano-D\'{\i}az et al. in preparation).

\section{Lopsidedness} \label{Lopsidedness}

Next we present a quantitative analysis of the asymmetry properties of NGC 3367 in various wavelengths. We start with the analysis
of the optical and NIR light by means of the Concentration-Asymmetry-Clumpiness parameters. These three structural 
and morphological indexes constitute the so-called $CAS$ system, which has been proposed for distinguishing galaxies at different stages 
of evolution (Conselice 2003, hereafter C03 and references therein). Briefly, the concentration index $C$ is defined as the ratio of the 
80\% to 20\% curve of growth radii (r80, r20) within 1.5 times the Petrosian inverted radius at r($\eta$ = 0.2) $r$ normalized by a 
logarithm: $C = 5 log(r80/r20)$. The concentration of light is related to the galaxy light (or stellar mass) distribution.  

The asymmetry index is the number computed when a galaxy is rotated $180^{\circ}$ from its center and then subtracted from its prerotated 
image. The summation of the intensities of the absolute-value residuals of this subtraction are compared with the original galaxy flux. 
This parameter is also measured within $r$. The $A$-index is sensitive to any feature that produces asymmetric light distributions. This 
includes galaxy interactions and mergers, large star-forming regions, and dust lanes.  

Galaxies undergoing $SF$ are very patchy and contain large amounts of light at high spatial frequency. To quantify this, the clumpiness 
index $S$ is defined as the ratio of the amount of light contained in high-frequency structures to the total amount of light in the 
galaxy within $r$. The $S$-parameter, because of its morphological nature, is sensitive to dust lanes and inclination.

\subsection{$CAS$ Structural Parameters}

Figure 7 shows the wavelength-dependent position of NGC 3367 in the $CAS$ parameter space 
($UBVRIJK$ symbols). The loci (average and its standard deviation) for a sample of isolated Sa-Sb and Sbc-Sm galaxies 
(Hern\'andez-Toledo et al. 2007, 2008) in the $R-$ band $CAS$ planes are indicated as crosses and continuous error bars. 
The solid boxes indicate the maximum amplitude of variation (lower and upper limits) of the $CAS$ parameters from $B$ to $K$ bands 
for the isolated disk sample. For comparison, the $R-$band $CAS$ averages and standard deviations of galaxies in interacting S+S pairs 
(open circles; Hern\'andez-Toledo et al. 2005), Starburst (long-dash) and Ultra Luminous Infrared (ULIR) galaxies (short-dash;C03) 
are also plotted.

\placefigure{fig-7}

NGC 3367 experiences a strong migration in the $CAS$ diagram. While the $UBVR$ band $CAS$ values place it as a typical Starburst, 
the $R$ band asymmetry ($A(R) = 0.35$) is also consistent with that of weakly interacting galaxies (Hern\'andez-Toledo et al. 2005) and 
at the end its locus in the $JK$ bands place it within the boundaries of nearly isolated galaxies. The Population I disk of young stars 
(and dust) yield a significant contribution to the optical $U, B, V$ $CAS$ distributions, making their average values largely decoupled 
from the corresponding to older $J, K$ stars. The clumpiness parameter $S$, compares the amount of light in 
star-forming clusters and young associations to the light in a more diffuse older disk population. This measure correlates with Hydrogen 
recombination lines (H$\alpha$) and gives an indication of recent star-formation activity. We estimate the star formation rate (SFR) in NGC 3367 
through $L_{FIR}$ (see Table 2). $SFR_{FIR}(M_{\odot} yr^{-1})$ = $1.40 \times 10^{-10} L_{FIR}(L_{\odot})$ (Devereux \& Young 1991) holds for the 
high-mass end ($M \geq 10 M_{\odot}$) and was corrected by the factor given in Devereux \& Young (1990) to take into account the fact that 
$L_{FIR}$ is used instead of $L_{IR}$. A SFR $\sim 3.1 M_{\odot} yr^{-1}$ in NGC 3367 (compared with 5 $M_{\odot} yr^{-1}$ reported 
by Garc\'{\i}a-Barreto et a. 2005) is an order of magnitude higher than the median value SFR $\sim 
0.27 M_{\odot} yr^{-1}$  obtained for a sample of isolated galaxies of similar (Sc) morphological types (Hern\'andez-Toledo et al. 2001), consistent 
with NGC 3367 being in the Starburst region of the $CAS$ diagrams.

\subsection{NIR Lopsidedness (Fourier Analysis)}

The light distribution in many galaxy disks is non-axisymmetric or lopsided with a spatial extent much larger along one half of a galaxy 
than the other. Nearly 30 $\%$ of galaxies have significant lopsidedness in stellar disks  with an amplitude of $>$ 10$\%$ measured as 
the Fourier amplitude of the $m=1$ component normalized to the average value (Rix \& Zaritsky 1995, Zaritsky \& Rix 1997, Bournaud et 
al. 2005).   

We calculated the azimuthally averaged $m=0, 1$ and $2$ Fourier amplitude profiles ($A_0$, $A_1$ and $A_2$) in NGC 3367, following the prescription 
described by Rix \& Zaritsky (1995) in the NIR light. The surface brightness distribution $\mu(R,\phi)$ can be expressed as a Fourier 
series:
\begin{eqnarray}
\mu(R,\phi)/<\mu(R)> =\sum^{\infty}_{m=1}A_m(R)e^{im[\phi-\phi_m(R)]}
\end{eqnarray}

\noindent where $\phi$ denotes the azimuthal angle, $m$ the azimuthal number, and $A_m$ and $\phi_m(R)$ are the associated Fourier 
amplitude and phase, respectively. The isophotal surface brightness at radius $R$ is given by $<\mu(R)>$.

Each filter image was re-binned onto a $(R,\phi)$ grid, using 30 bins in radius and 24 bins in azimuth. We have fixed a minimum radius 
of 4 pixels and  a maximum radius of 60 pixels. Notice that we quote the analysis in the NIR images to 60 pixels 
corresponding to a maximum radius of 51 arcsec, shorter than those in the optical images, but still ensuring enough signal in the 
disk emission at the $J$, $H$ and $K$ bands. 
   
Figure 8 shows the results of $A_m(R)/A_0(R)$ for the $m =1-2$ modes in the $J$ band. Radius is plotted in pixel units (1 scale-length 
$=$ 22.54 pix or 19.16 arcsec). The scale-length was calculated by fitting a straight line in the disk domain in the 
$A_0(R)$ Fourier amplitude (upper panel; which formally represent the mean light profile) consistent with the values estimated from our 
former surface brightness profile fittings. The intermediate and lower panels show the $m=1$ and $m=2$ amplitudes and their corresponding 
phases, respectively.

\placefigure{fig-8}

As can be seen, the NIR observations enclose light to almost $3 R_d$. The results for $<A_1>$ are certainly different when we use 
different regions to calculate the mean. For example, in the inner disk, up to 1.0 $R_d$, the $<A_1>$ distortion is weak (0.05 
or lower), similar to that observed in other galaxies (c.f. Rix \& Zaritsky 1995). For radii between (1.5$R_d$, 2.5$R_d$), $<A_1> = 
(0.12, 0.11, 0.11)$ for the $J$, $H$ and $K$ bands respectively, close to the mean value ($<A_1> = 0.10$) in the sample of 149 
mostly field galaxies in Bournaud et al. (2005). More specifically, between 1.0$R_d$ and 2.0$R_d$ there is a local maximum at 
0.15 level that corresponds to a prominent knotty region near the northeastern end of the bar, a value that is higher than the 
average value for field galaxies. The gradual change in phase in that region denotes that such feature lies along an arm winding 
upward to the north.  Since the outer disk in the $J$ band is weakly detected, to visualize the behavior of $<A_1>$ in the boundary 
of 60 pix and beyond, we use a deep $R$ band image instead, obtaining that $<A_1>$ appears to decrease below 0.05 level, 
with a roughly constant phase. By comparison, the phase is nearly constant with radius in the data of Rix \& Zaritsky (1995). 
NGC 3367 could be regarded as slightly non-symmetric, at the nominal intermediate radii. In the lower panel of Figure 8, the 
$m=2$ amplitude and phase confirms the existence of the bar within 1.0$R_d$. Between 1.0$R_d$ and 2.0$R_d$ there is another 
local maximum of similar amplitude, that according to the associated phase, corresponds to the inner arms enclosing the bar.

\subsection{H$\alpha$ Lopsidedness}

Figure 9 shows the results of the H$\alpha$ Fourier $A_m(R)/A_0(R)$ $m =1-2$ amplitude and phases estimated from 
an image resulting after collapsing the 48 channels of a Fabry-Perot H$\alpha$ data cube (Garc\'{\i}a-Barreto \& 
Rosado 2001).

\placefigure{fig-9}

The asymmetry in H$\alpha$ is presented at about the same radial distance as that in near-IR study (1 pix $=$ 0.6 arcsec).  
The two tracers show qualitatively similar features in the $m=1,2$ modes, except for the significantly higher amplitude observed 
in the H$\alpha$ tracer. In the 80 pix ($\sim 50 ~arcsec$) region the H$\alpha$ $m=1$ phase 
reveals the prominence of the N-W-S outline that dominates the global H$\alpha$ appearance of 
NGC 3367 as devised in the left panel of Figure 9. This is on line with the results of Jog (1997), who claims that in a 
galactic disk, a lopsided distribution should induce an azimuthal asymmetry in star formation. For the $m=2$ mode, similar 
to the $m=1$ case, a significant increase in amplitude is also observed. NGC 3367 is locally and globally more 
asymmetric in the H$\alpha$ light than in the NIR wavelengths.

\subsection{HI Lopsidedness}

The frequency of asymmetries among spiral galaxies has alternatively been estimated from the global HI profiles of a large 
sample of field galaxies (Richter and Sancisi 1994). Examples of asymmetric global HI profiles can be found in M101 and NGC 4395. 
About 20\% of the systems examined showed strong asymmetries and up to more than 50\% of the whole sample showed some mild 
asymmetries. That result has been confirmed by a 21-cm HI survey of 104 isolated galaxies with the Green Bank 43-m telescope 
(Haynes et al. 1998) and also by more recent observations which image the HI distribution and the kinematics of a large sample 
of galaxies (Westerbork H I Survey of Spiral and Irregular Galaxies; WHISP). At least one half of nearly  300 objects from 
WHISP shows some lopsidedness either in the HI distribution, in the kinematics or in both. It should 
be noted that these lopsided galaxies seem to be in non-interacting systems and that, therefore, the lopsidedness cannot be 
explained as a present tidal effect.  In this paper we started the basic analysis of NGC 3367 HI 
asymmetry properties. The upper panel of Figure 10 shows a grey scale HI intensity map obtained from the moment 0 VLA data cube 
(natural weighting) showing various contours above the sigma level. The lower panel shows the corresponding integrated HI velocity 
profile.   

\placefigure{fig-10}

The moment 0 intensity map shows (i) an HI hole in the central region, mostly covered with molecular gas (Garc\'{\i}a-Barreto et al. 
2005) and (ii) an HI asymmetry in the column density with higher values at the northwest at good statistical significance. The 
shape of the global HI column density distribution when superposed on an optical image, strongly emphasizes that apparent truncation 
viewed not with the same abruptness in the western than in the northeastern side in the $UBV$ and H$\alpha$ images. The integrated 
VLA profile shows higher flux density in the redshifted side of the spectra, consistent with the available single-dish HI velocity 
profiles for NGC 3367 in the literature. The absence of a drop-off in the integrated emission line at the galaxy systemic velocity 
is consistent with the excess column density in the northwest minor axis region (P.A. major axis $\sim 51^{\circ}$).

To estimate the HI asymmetry, we first computed the flux ratio $f = F1/F2$ between the two halves of the profile. We have taken the 
total width at 20\% of the peak flux density, and estimated both the extremities of the profile and its center. The asymmetry parameter 
is defined as $A = 10(1 - F1/F2)$ yielding a value $A = 2.2$ for NGC 3367. For comparison, notice that in the sample of moderately 
isolated galaxies in Bournaud et al. (2005) only 26\% (18 out of 80 galaxies) have $A > 2$. However, in order to characterize possible 
instrumental or telescope effects that can artificially increase intrinsic HI asymmetries, we estimated the HI lopsidedness in NGC 3367 
from integrated spectra retrieved from other single-dish telescopes (Huchtmeier \& Seiradakis 1985 [Effelsberg]; Mirabel \& Sanders 1988 
[Arecibo]; Staveley-Smith \& Davies 1988 [Jodrell-Bank]; Springob et al. 2005 [Arecibo]), obtaining a mean value of  $A = 1.8$ with a 
scatter of 0.7. Notice that even the lowest estimated value is similar to the average asymmetry (stellar component) reported by 
Bournaud et al. (2005) for galaxies in the local universe, indicating that in NGC 3367 HI asymmetry is detected with good statistical 
significance, regardless of the instrumentation.

We also explored the VLA HI velocity channels. However at the current signal level we do not find convincing evidence of neutral gas 
fall or ejection. The fact that we find a significant lopsidedness in the HI gas component in NGC 3367 suggest that the triggering event 
must have happened some time in the past. In section 8 we use gas relaxation time arguments in order to constrain the time of a perturbation 
event.

\section{Low Surface Brightness Structures and Companions} 
\label{Isignal}

The most natural explanation for NGC 3367 perturbed morphology and asymmetries is a recent encounter with a low-mass galaxy.  
A deep 7.2 $\times$ 7.2 arcmin $R$ band image of NGC 3367 has been produced in order to look for low surface brightness features 
reminiscent of possible tidal interactions. The final image in Figure 11 is the sum of one 1800 seconds exposure, two 1200 seconds 
exposures, and one 900 seconds exposure and it is presented in logarithmic scale. The circular contour illustrates the position 
of regions with S/N $=$ 3.

\placefigure{fig-11}

This image highlights the more circular nature of the disk and some possible diffuse light structures at the outskirts. To quote 
the depth of the image we built a pixel representation of two images dubbed as the variance and the average by combining all the 
images intervening in Figure 11.  Since the calibration of this deep image is not currently available, we relied in our calibrated 
short exposure images (typically 15 min) at the same band, as shown in Figure 2. We estimate a S/N ratio $\sim 3$ at a maximum 
radius of the order of 80 arcsec (see circular contour), with a detection limit at that position $\mu_R \sim 26.0 ~mag 
~arcsec^{-2}$. That level is of the order of that in recent studies looking for similar features in isolated galaxies (Smirnova et al. 
2010). We also applied histogram equalization and unsharp masking transformations to the image similar to recent studies 
(Mart{\'{\i}}nez-Delgado et al. 2010), however because of the low S/N beyond 80 arcsec, we find no reliable evidence for the 
existence of external low surface brightness features at the current depth of this image. A careful control of the background 
fluctuations due to flat-field residuals, internal reflections, ghosts and scattered light that were not considered in detail 
in the present image, deserves special attention. Deeper and careful observations are still necessary to provide unambiguous 
evidence of any tidal structure at the outskirts of NGC 3367 (Mart{\'{\i}}nez-Delgado et al. 2010).

A procedure, coined as blinking colors, was implemented to search for small companions within the disk 
and/or in the close neighborhood of NGC 3367. A fact is used that under some circumstances, a small galaxy laying in front 
of a bigger one could still be recognized if its own colors (stellar population, however dust effects should be accounted) are 
different enough from those in the background galaxy. To that purpose, we have produced various combinations of color index maps, 
namely $U-B$, $B-V$, $U-K$, $B-K$, from our $U$-to-$K$ band observations blinking the parent images along with their corresponding 
color maps with the hope of visually identify any feature not sharing the observed colors of the ground host galaxy. We also 
produced a pixel representation of each color map to have a more quantitative visualization of the color properties. Unfortunately 
these procedures produced no positive detections. Figure 12 (left panel) shows a ($B-V$) color-index map and the corresponding 
distribution of pixels (right panel) within a radius of 60 pixels, well within the disk and assuring a pixel S/N ratio of 3. 
Pixels were binned to the order of the seeing and then corrected for galactic extinction. Blue color 
represents values (0 $<$ ($B-V$) $<$ 0.5), pixels in green (0.5 $<$ ($B-V$) $<$ 0.75), pixels in yellow (0.75 $<$ ($B-V$) $<$ 1.0) and 
pixels in red ($B-V$) $>$ 1.0.

The color pixel representation suggests the existence of structure adjacent to the compact nucleus in the (0.75 $<$ ($B-V$) $<$ 1.0) 
color range (central green-yellow-red transition region in the upper right panel) resembling a circumnuclear 
component. Whether there is a circumnuclear ring or disc is currently being assessed using archive Spitzer images and CO kinematics 
and will be presented elsewhere (see also the residual image after our 2D bulge/bar/disk decomposition in Figure 3). 

Finally, we mention our search for H$\alpha$ emission satellite galaxies in the close neighborhood of 
NGC 3367. A series of narrow-band ($\Delta \lambda \sim 80 \AA$) H$\alpha$ images with a field of view of 4.5 $\times$ 4.5 arcmin 
were obtained at the OAN-SPM 1.5m telescope with a set of extragalactic H$\alpha$ filters tuned at different red-shifts (covering a 
radial velocity range from 1000 ~km ~$s^{-1}$ up to 6000 ~km ~$s^{-1}$) and sharing a degree of velocity overlap among adjacent 
filters. A blinking of the images at different recession velocities with respect to that in the Gunn $r$ broad-band yielded no 
positive identifications. Recently Ivory \& James (2011) showed that this kind of searches reveal star forming satellite galaxies. 
 
In the mean time, and with the above evidence, NGC 3367 is tentatively added to the list of candidate spirals showing various signs 
of perturbation, but no large nearby companions, these systems may have undergone  some kind of smoothlike mass accretion
during the past Gyrs.

\placefigure{fig-12}

\section{Discussion}

\subsection {Evidence for Recent  NGC 3367 interaction with a small  galaxy?}

An intuitive explanation for NGC 3367 optical asymmetric morphology and large bipolar synchrotron emission might be a recent merger 
with a small galaxy since NGC 3367 has no large optical companion closer than 500 kpc (Garc\'{\i}a -Barreto, Carrillo and Vera-Villamizar 
2003). A gas rich galaxy may be a suited candidate because the asymmetric star formation distribution. NGC 5548 is an example of 
another Seyfert galaxy that is not suffering a major encounter, but that presents faint stellar structures remnant of past interactions.  
Other important differences in this comparison are the galaxy morphological type and the high density environment in NGC 5548. Hence, 
finding evidence for a past possible merger event in NGC 3367 is important in order to define a plausible scenario that explains its 
properties. A multiwavelength search was performed in order to look for a possible remnant signature or a dwarf perturber, within the 
disk and in its close neighborhood. Initially we used our data at various optical $H\alpha$ and infrared wavelengths and different 
enhancing procedures, as presented in section\ref{Isignal}. So far no convincing evidence was found, thus placing constraints to the time of 
any possible galaxy accretion event.

Different studies have analyzed galaxy disk response to minor mergers. For example, Kazantzidis et al. (2008) have recently 
studied the signature of galaxy merger histories in the context of $\Lambda$CDM scenarios  mostly focusing in gravitational interactions 
between existing thin galactic disks and small galaxies with masses comparable to the LMC. They conclude that if a merging event 
occurs, it could at least, be responsible for the formation of thick disks, central bars, low surface brightness ring-like 
configurations and faint filamentary structures around a main galaxy, like in NGC 3367 making this scenario compelling. Unfortunately 
Kazantzidis et al. (2008) did not show the perturber remnant distribution. In the following sections we compare instead 
with similar experiments by Villalobos \& Helmi (2008) because they present both the perturber satellite and the host disk 
distributions across different times. At the current state of our  photometric search we can not support the idea of a 
recent minor merger with an LMC-like galaxy; at most we can constraint the event to the distant past. Alternatively, if we consider a 
recent interaction with a dwarf galaxy, its mass should be considerably lighter than NGC 3367 disk, otherwise narrow stellar 
streams may still be detectable.

\placefigure{fig-13}

\subsection{Possible Origin of Disk Lopsidedness}

NGC 3367 asymmetry properties were characterized using different techniques and different wavelengths. The  high redshift motivated 
asymmetry parameter in the $CAS$ diagram goes from a strongly perturbed galaxy values and a strong Starburst one in  $U$ and $B$ bands 
(young stellar component and dust) to the ones corresponding to a nearly isolated spiral galaxy in the $J$ and $K$ bands 
(average stellar population). 

The azimuthally averaged $m=0, $1 and $2$ Fourier amplitude and phases, ($A_0$,$A_1$,$A_2$ and $\phi_1$, $\phi_2$ and $\phi_3$ ) in 
the $K$ band (also the ones from our deep $R$ band image for the external regions) were estimated finding that:
 
 (i) Up to  1.0 $R_d$, the $<A_1>$ distortion is weak (0.05 or lower).
 
 (ii) between 1.0$R_d$ and 2.0$R_d$, $<A_1>$ is  slightly higher than the average for normal galaxies at the same radius, reaching 
$<A_1>$ $\sim$ 0.15 and
 
 (iii)  for radii greater than  2.0$R_d$, $<A_1>$  approach a low limiting value of 0.05.

We also estimated the Fourier modes amplitude and phase profiles using the $H\alpha$ intensity map finding that the amplitude ranges 
at least a factor of 2 higher than in the old stellar component, in the same radial interval. Finally, as shown 
in section 6, the HI integrated line profile and HI surface density distributions are highly asymmetric, indeed more than the average 
stellar component. 

The diagnostics discussed above indicate that the $A_{1}$ asymmetry is present across wavelengths and also in the ionized and 
neutral gaseous component, suggesting that lopsidedness is a property of the potential and not only an artifact of the SF pattern. 
Several scenarios have been suggested as possible origin for lopsidedness,  from tidal perturbations triggered by
massive companions, satellite galaxy accretion, high-speed  encounters (flyby), gas accretion through cosmic filaments, and internal 
instabilities (Rix \& Zaritsky 1995, Walker et al. 1996, Bournaud et al. 2005, Dury et al. 2008).  The isolated galaxy instability 
is an attractive possibility based on NGC 3367 environment, however, it is not obvious why different mass components show different
lopsidedness amplitude.  Among the external perturber scenarios, Bournaud et al. 
(2005) extensively discuss the survival time for the asymmetry triggered inside the disk by a satellite galaxy accretion.  A 
comparison of our $A_{1}$ mode analysis with their figure 12 could be used to a constrain the time elapsed after a possible minor 
intruder to be longer than 2 Gyr.  

Extra constraints can be set based on the lack of a central classical bulge, the absence of stellar low surface brightness 
plumes at the disk edge (at the limit of our observations) as well as the lack of localized regions with different stellar 
population (colors).  Our finding of a pseudo-bulge in NGC 3367 suggests that the stellar and total mass of a past perturber was 
small enough to avoid the formation of a classical bulge. We use the study presented recently by Villalobos and Helmi (2008) in order 
to give an interpretation to the absence of low surface brightness plumes. This study show N-body simulations aimed to track the 
formation of thick discs at different epochs, conveniently they analyze the evolution of the host disk and satellite perturber 
stellar distribution. They included satellite galaxies having total masses 10-20\% that of the host galaxy disk, 
self-consistently modeled as a stellar component immersed in a DM halo. The stellar components have either a spheroidal or disky stellar 
structure. Particularly interesting are the low surface brightness shells, especially visible in the outskirts of the final thick 
discs, that last for about 1.5 to 2 Gyr after the merger has been completed. Diagnosing the existence of these shells acquires relevance 
in the case of NGC 3367. If we compare with figure 7 in Villalobos and Helmi (2008), we see that there are not noticeable thin stellar 
structures after 4 Gyr providing a lower limit for the accretion time if an LMC type accretion event happened in NGC 3367. If we 
consider lower mass satellites like dark subhalos containing gas, the time constraint will be definitively shorter (Chakrabarti \& Blitz 2010). 
However even this scenario does not completely explain why the gaseous component, whose relaxation time is way shorter than the 
stellar one, still presents important asymmetries both in H-alpha and HI, unless they are currently perturbed.

The lower panel in Figure 7 shows two solid boxes indicating the maximum observed variation in clumpiness from $B$ to $K$ bands 
for a sample of isolated galaxies of Sa-Sb and Sbc-Sm types. NGC 3367 is clearly more clumpy in the optical $UBV$ bands   
than isolated galaxies of similar morphological types. This large abundance of star formation knots (that are appreciated not only along 
the spiral arms in Figure 1) provide complementary hints pointing to a past gas accretion scenario.

The accretion of cold gas may lower the Toomre local stability parameter ($Q$) triggering the 
clump formation and enhancing a bar (Sellwood \& Moore 1999; Block et al. 2002). It might also produce asymmetrically distributed 
episodes of star formation, producing a lopsided structure particularly in the young stellar component and gas (Bournaud et al. 2005). 
The same mechanism can also explain the Starburst phase (Dekel et al. 2008) and the high bar amplitude for the morphological type. A combined 
mechanism of external accretion and bar radial transport of gas and stars might be responsible of the color profiles presented in 
section \ref{sec:profiles}, and also of the large $A_1$ amplitude in blue colors (Mapelli et al. 2008).  Although gas accretion seems an 
attractive scenario, from the HI VLA observations we find no convincing evidence for extraplanar gas in the gas kinematics, concluding 
that the gas asymmetry may be the result of a past event.  The gas relaxation time constraints the time of the event to be a few million 
years. As a coincidence, if we assume that the radio jet speed is one-tenth the speed of light with a projected radius of 6.5 kpc, 
the ejection time is constrained to a similar short time scale suggesting that the same accretion event may be responsible of both 
morphology and nuclear activity.

 \subsection{Evidence of Secular Evolution in NGC 3367}  
 
Bar structural and kinematic properties are correlated with its dynamical history. We determined the bar strength using 
two complementary methods, showing that NGC 3367 has a strong bar. The $B-I$ color map  revealed conspicuous dust lanes whose shape is 
sensitive to the bar pattern speed. Specifically, the  dust lane orientation non-parallel to the bar major axis may suggest a relatively 
slow bar, which is not unexpected in late-type galaxies (Salo et al. 1999). However a definitive answer must wait to direct estimation.
The bar ellipticity and overall structure suggest a  spheroid presence absorbing angular momentum from the bar triggering its growth. 
Because the 2-D photometry does not indicate an important bulge, the bar structure may indicate the presence of a dark matter component. 
However, the lopsided structure and the possible gas accretion introduce an uncertainty  that can not be sorted out without more 
investigation of the angular momentum history if m=2 and m=1 modes coexist in gas accreting disk galaxies. Comparison of images in $K$ 
and $U$ and $B$ bands, suggest that secular evolution has triggered gas inflow and star formation along the bar. In particular the $U$ and $B$ 
morphology suggest a very thin and curved bar and also a blue compact nucleus, suggesting star formation along the bar.  The 
$NIR$ torque strength parameter indicates that NGC 3367 has an outstanding strong bar, this probably relates to the nuclear activity and 
is consistent with the molecular gas abundance in the bar region. 

The bulge of NGC 3367 is exponential, similar to the so-called ''pseudobulges''. Bulges of this type 
are thought to be generated via internal secular evolution processes (see, e.g., Kormendy \& Kennicutt 2004). Given that NGC 3367 is isolated, 
secular evolution is likely the dominant evolutionary mechanism. Pseudobulges typically resemble small discs embedded in larger ones or 
flattened spheroids. The strong bar presence and the large abundance of young stellar clumps in the disk (Garc\'{\i}a-Barreto et al. 2007; 
Elmegreen et al. 2008), support this scenario as a possible one for NGC 3367 bulge formation. Foyle et al. (2009) observationally estimated 
the effect of gravitational torques due to the present day global disk stellar distribution. They discussed whether these torques are 
efficient at transporting angular momentum within a Hubble Time. They found that torques due to the stellar disk lead on average to 
outward angular momentum transport over much of the disk (r $\leq 3 ~r_{exp}$). Gravitational torques induce angular momentum change on 
time scales shorter than a Hubble Time inside one scalelength, where they act on a time scales of 4 Gyr. If we compare with Foyle et al. 
(2009) results, secular evolution is thus expected to be effective in the inner parts of NGC 3367 and perhaps is also connected with the 
nuclear activity, however, the details for this process and its relationship to the fueling of the central active galactic nucleus (AGN) 
in NGC 3367 need to be worked out.

\subsection{Nuclear Activity and NGC 3367 Structure}

NGC 3367 is a Seyfert2/Liner galaxy presenting a radio continuum jet, however it is still considered a radio quiet galaxy.  A natural 
question is what is the connection, if any,  between the nuclear activity and the asymmetric galaxy morphology. As we concluded above, 
the most compelling candidate scenario for being responsible for NGC 3367 morphology is a past gas rich accretion event.  
Currently cosmological galaxy formation scenarios provide an explanation of both possibilities; galaxies smaller than $10^{12} M_{\sun}$ 
receive significant accretion (small galaxies and gas) through cosmic filaments (Dekel et al. 2009). The main argument favoring this scenario 
is the lack of dynamically young stellar signatures associated to a recent tidal event.  Specifically the  absence of thin tidal tails 
constrains any interaction event to some dynamical times in the past (3-5 Gyr) (c.f. Villalobos \& Helmi 2008),  however the gas 
component that has a shorter relaxation time shows both a global ($A_1$ mode) and localized overdensities. That short gas relaxation 
time compared with the stellar component, suggest a perturbing event in the last hundreds of million years. The images in short 
($U$ and $B$) wavelengths reveal young stellar population all the way along the bar up to the nucleus suggesting that a continuous gas supply 
has been channeled by the bar towards the nuclei.

Cosmological simulations of galaxy formation show that accreted gas (either from satellites or cold flows) frequently has a different 
angular momentum orientation compared with the existing disks originating warps or polar disks (Maccio et al. 2007, Roskar et al. 2010). 
  The highly inclined jet in 
NGC 3367 (Garc\'{\i}a-Barreto et al. 2001) might well be the result of a nuclear inclined disk triggered by misaligned gas accretion. 
The fact that statistical studies show an unexpectedly high fraction (22 $\%$) of active isolated galaxies (Sabater et al. 2010) as 
well as asymmetric isolated disk galaxies (30$\%$) (Hern\'andez-Toledo et al. in preparation), and also that roughly half ($44\%$) of the 
Seyfert/LINER galaxies in a distance limited sample show radio jet-like structures at kpc-scale (Gallimore et al. 2006), 
suggest that asymmetric gas accretion, isolated lopsided, barred galaxies and nuclear activity may be correlated.  If this is 
confirmed by the future corresponding statistical multiwavelength studies, large scale gas accretion may raise as an important 
source of nuclear galaxy activity. Furthermore, we can expect that at high redshift, when filament accretion rate was more 
vigorous and common (gas and satellites), the AGN feedback triggered only by the cold accretion mode and enhanced by disk asymmetries 
may be considerably more important than typically assumed. 

NGC 3367 can be added to the list of local late-type galaxies that lack a classical bulge but that host an obscured AGN (McAlpine et al. 2011; 
Ghosh et al. 2008; Desroches \& Ho 2009). This type of galaxies may reveal kpc-scale jet-like structures in deep enough radio imaging of local 
surveys. The black hole mass in this pseudo-bulge galaxy has been recently estimated using various 
methods, ranging between $10^{5}-10^{7}M_{\sun}$ (McAlpine et al. 2011).

 \subsection{Constraining  Formation Scenarios}   
The discussion presented above favors the following facts:   

(i) NGC 3367 is a galaxy likely localized in a low density environment, the nearest dense structure is the Leo Group, however the 
difference in line of sight velocity is  near to 2000 ~km ~$s^{-1}$, considerably larger than Leo escape velocity, therefore we think 
they are unlikely  bounded.

(ii) The importance of major mergers in NGC 3367 assembling  and even of minor mergers with massive satellites is severely 
constrained because of the lack of a classical bulge as well as the absence of bright and thin stellar streams. In fact the B/D 
ratio 0.07-0.11 marginally favors a pseudobulge, perhaps caught in the act of formation. This galaxy is an example of the late type 
active ones discussed by Schawinski et al. (2011), the possibility of a gas rich minor merger although not discarded, is constrained 
to current very low mass satellites or toward the past 3-4 Gyr. For comparison, typical fractional lopsidedness amplitude $A_1$ within 
the central 5 kpc from 2MASS images of advanced mergers of galaxies (Jog \& Maybhate 2006) go from $\sim$ 0.12 to 0.2, higher than the 
values found in the inner regions of NGC 3367. Notice that those advanced mergers were selected as having merged into single nucleus but 
still show indications of interactions, including visible tidal tails, contrary to what is observed in NGC 3367.  

(iii) The stronger asymmetry found in the young stellar component and gas constrains a possible perturbation to the gas relaxation time 
scales which are shorter than the stellar relaxation times (Gyr scale), therefore the gas perturbation must has happened a few 
million years in the past.

(iv) The high clumpiness parameter value and the global Starburst  phase (see Figure 7) may be explained if the gas richness is high.
Based on the integrated HI line profile and CO observations, NGC 3367 has a gaseous mass of a few $10^{10} M_{\sun}$ (Huchtmeier 
and Richter 1989; Garc\'{\i}a -Barreto et al. 2005). A naive estimation of the stellar mass is severely affected by the AGN light 
and by dust extinction. For example, if we use published $K$ band photometry, a maximum rotation velocity close to 300 ~km ~$s^{-1}$ 
or larger is inferred from the Tully-Fischer relation. In contrast, Garc\'{\i}a-Barreto \& Rosado (2001) present estimates of the 
galaxy maximum rotation velocity in the range of (195-225) ~km ~$s^{-1}$, not far from the reported values for the Milky Way. As a 
comparison, the Milky Way has a stellar mass around 6 $\times 10^{10} M_{\sun}$ and a gas to baryon fraction of 13\%. Avoiding a 
detailed modelling of the disk and nuclear emission required to estimate the stellar mass, and using recent theoretical studies 
(Baldry et a 2008, Rodr\'{\i}guez-Puebla et al. 2011) that suggest a relationship between the total galaxy mass and the stellar mass, 
we can assume a $V_{max}$ = 225 ~km ~$s^{-1}$ for NGC 3367 and infer a stellar mass of the order of 6 $\times 10^{10} M_{\sun}$. 
Since the gas mass can be directly estimated from the observations, the inferred gas to baryon fraction attains 16\%. On the 
other hand, if we assume the lower limit of $V_{max}$, we can infer a gas fraction of about (25-30)\%. The estimated gas richness 
in NGC 3367 (16-30\%) when compared versus galaxies of similar $V_{max}$ like the Milky Way (13\%), suggest that the gaseous 
disk has been experiencing fragmentation and localized stellar bursts (see Figure 7 and Garc\'ia-Barreto et al. 2007).  

(v) The scenario  resembles truly rich gas galaxies at $z > 1$, where disk fragmentation is believed to contribute to bulge 
formation (e.g. Ceverino et al. 2010). However, notice that NGC 3367 is less gaseous and less violently unstable. If this analogy 
is correct, disk clumps infall due to dynamical friction may be forming now a pseudobulge (Elmegreen et al. 2008) and may be feeding 
the central blackhole growth and activity. Recent observational evidence (Kormendy et al. 2011) found no correlation 
between pseudobulge and blackhole masses, setting a question mark on the importance of clump migration as a blackhole feeding 
mechanism.

\section{Conclusions} \label{S5}

From a set of optical, NIR and H$\alpha$ Fabry-Perot observations carried out at San Pedro M\'artir, National Optical Observatory in
M\'exico and from archive HI VLA data we studied some structural properties in NGC 3367 and tried to constrain possible origins for
its asymmetric structure.  
 
Our surface photometry analysis and coupled 1D-2D bulge/bar/disk decomposition procedures indicate that both the bulge and disk of 
NGC 3367 are consistent with a double exponential luminosity distribution in all the observed wavelengths with a rather    
small B/D luminosity ratio suggesting a R'SBcd(r) morphological type with a pseudobulge, constraining the mass of
possible past perturber to rather low values. From the position of the galaxy in the $CAS$  structural diagrams, it is inferred that 
NGC 3367 is in a global 
Starburst phase, with a SFR $\sim 3-4 M_{\sun} yr^{-1}$.   

A qualitative comparison of the observed dustlane morphology from ($B-I$) color index maps with the results from  
bar numerical simulations suggest that NGC 3367 may host a relatively slow bar. Bar torques, estimated from the NIR images (Buta \& Block 2001) 
yielded a bar strength $Q_b$ = 0.44, placing NGC 3367 as a galaxy with a very strong bar. The presence of such a bar may be behind the complete 
absence of HI gas in the central regions of NGC 3367, mostly covered with a high amount of CO gas instead. The absence of such bars 
may explain why other similarly asymmetric galaxies like M101 do not currently present strong nuclear activity.    

Global and local asymmetry in NGC 3367 has been analyzed from three alternative indicators: 1) through the $CAS$ structural parameters 
in the optical and NIR light, 2) from an estimate of the Fourier amplitude of the $m$-components of the density distributions in the 
NIR and H$\alpha$ light and finally, 3) from a new analysis of the HI VLA integrated profile and the moment 0 HI intensity distribution.

NGC 3367 shows a moderately high value of  $A_1$ in its intermediate stellar disk (0.15) but it is negligible in the inner and outer 
regions.  An asymmetry a factor of 2 higher is revealed in the H$\alpha$ light and integrated HI profile instead. This strong decoupling 
in asymmetry with respect to the mean stellar component challenge an internal disk instability origin for lopsidedness which seems natural 
based on NGC 3367 environment. However theoretical models have not considered gas and star formation yet (Dury et al. 2008), therefore 
further research is needed in order to assess such scenario that is relevant in our case.
  
Table 2 presents a summary of results inferred from the study presented here.

\placetable{tbl-2}

We performed a detailed search for dynamically young stellar signatures of a recent minor merger like stellar plumes or tidal 
tails, up to $\mu_R = 26 ~mag ~arcsec^{-2}$, similar to recent studies reporting companions in Seyfert and normal galaxies  previously 
considered as isolated (Smirnova et al. 2010, Mart\'{\i}nez-Delgado et al. 2010). We also searched for localized regions with different 
color distribution (stellar population). Both searches were unsuccessful but are useful to constrain either possible events like the 
incidence of minor mergers towards the past in such a way that any remnant is well mixed (3-5 Gyr) or to recent encounters with 
very small mass, even with dark matter satellites containing gas (Chakrabarti  \& Blitz 2010).

The strong decoupling in asymmetry of the HI integrated profile an the zero moment (intensity distribution) suggest that the gas 
which has a short relaxation time, has been recently perturbed mostly at the southeast and northwest, where we find some 
possibly related features but at low statistical significance. This picture and the highly inclined jet is consistent with the scenario 
of misaligned gas/satellite accretion by cosmic filaments as a trigger for lopsidedness. We suggest that the nuclear activity in this 
almost bulgeless late-type galaxy is also probably stimulated by the very strong bar. We speculate that at high red-shifts when 
gas/satellite accretion by filaments is more vigorous, the AGN feedback that is not triggered by major mergers and is enhanced by 
disk instabilities is likely more important than it has been typically considered. Statistical multiwavelength studies may verify this 
statement for local AGNs and future surveys will extend the analysis for high redshift.

\begin{acknowledgements}

HMHT, IP and JAGB thank the staff of the Observatorio Astron\'omico Nacional for the 
help in the observations. HMHT acknowledges support from grant
CONACyT-42810.  JAGB acknowledges partial support from PAPIIT (UNAM) grants IN-107806 and IN-112408.
OV acknowledges support from the PAPIIT (UNAM) grant IN-118108. IP acknowledges support from CONACyT.
HBA thanks E. Brinks for assistance to calibrate VLA HI-data.
 
\end{acknowledgements}

\clearpage

\begin{deluxetable}{ccccccc} 
\tablecolumns{8}
\tablewidth{0pc} 
\tablecaption{The $J$, $H$ and $K$ band bar strength in NGC 3367 \label{tbl-1}}
\tablehead{
\colhead{} Band & $<Q_T^{max}(R)>$ & }  
\startdata 
$J$  & 0.43 $\pm$ 0.05 &   \\
$H$  & 0.44 $\pm$ 0.06 &   \\
$K$  & 0.44 $\pm$ 0.09 &   \\
\enddata
\end{deluxetable}

\clearpage

\begin{deluxetable}{lllll}
\tabletypesize{\scriptsize}
%\rotate
\tablecaption{Morphological Properties of NGC 3367 in This Work and Other Properties. \label{tbl-2}}
\tablehead{
\colhead{}  } 

\startdata 

\multicolumn{1}{l}{} & Morphology &\\
\colhead{}{} \\
R'SBcd(r) & Circumnuclear disk or ring & \\
Incl $25^{\circ}$ $\pm 5^{\circ}$ at 80 arcsec
Bulge S\'ersic n = 1-1.8 \\
\hline

\multicolumn{1}{c}{} & Bar &\\
\colhead{}{}\\

length 17 arcsec (7 kpc) & P.A. $66^{\circ} \pm 4$ & \\
$\epsilon_{max}$ = 0.58 $\pm$ 0.05 
$<Q_T^{max}(R)>$ = 0.44 ($h_z = 325 pc$) \\
\hline

\multicolumn{1}{l}{} & Lopsidedness &\\
\colhead{}{}\\

$CAS$ $UBV$ Starburst & $CAS$ $R$ Weakly Interacting & \\
$CAS$ $JK$ Isolated galaxy & \\
Fourier $NIR$ & $\sim 0.05$ r $<$ 1.0 $R_d$ & \\
(0.12,0.11,0.11) ($J,H,K$) 1.5$R_d$ $<$ r $<$ 2.5$R_d$ &  
$\sim 0.05$  r$>$ 2.5$R_d$ \\
Fourier H$\alpha$ & $>$ 2 at all radii w.r.t. $NIR$ \\
Fourier HI(VLA) & $A = 2.2$ $>$ w.r.t. $NIR$\\
\hline

\multicolumn{1}{l}{} & Other Relevant Data &\\
\colhead{} \\

$(B-V)_T^{o}$                        & 0.47  & Garc\'{\i}a-Barreto et al. 2007  \\
$M_{HI} (10^{9} M\odot$)             & 7     & Hutchmeier \& Seiridakis 1985    \\
$M_{H2} (10^{9} M\odot$)             & $>$ 3     & Garc\'{\i}a-Barreto et al. 2005 \\
Log ($L_{B}$)                        & 10.68 & Tully 1998.                      \\
Log ($L_{FIR}) (10^{10} L_{\odot}$)  & 10.33 &                                  \\
SFR ($M_{\odot} yr^{-1}$)              & 3-4 &                                  \\

\enddata
\end{deluxetable}

\clearpage

\begin{figure}
\plotone{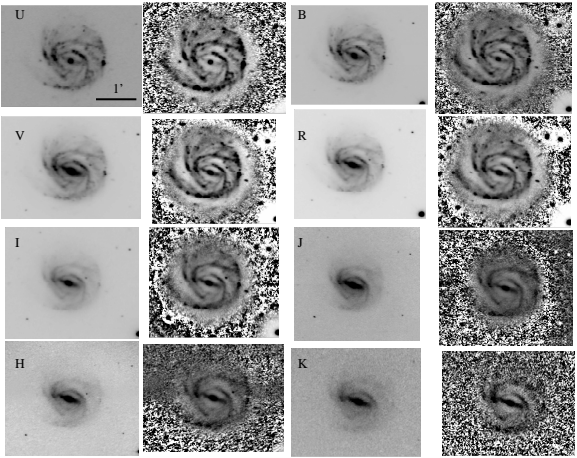}
\caption{$U$ to $K$ band images of NGC 3367 and their corresponding filtered-enhanced version (adjacent right-side images). 
The scale of the images is illustrated in the $U$ band image. All images are oriented according to the astronomical convention;
North to the top and East to the left.}
\label{figura1}
\end{figure}

%\clearpage

\begin{figure}
\plotone{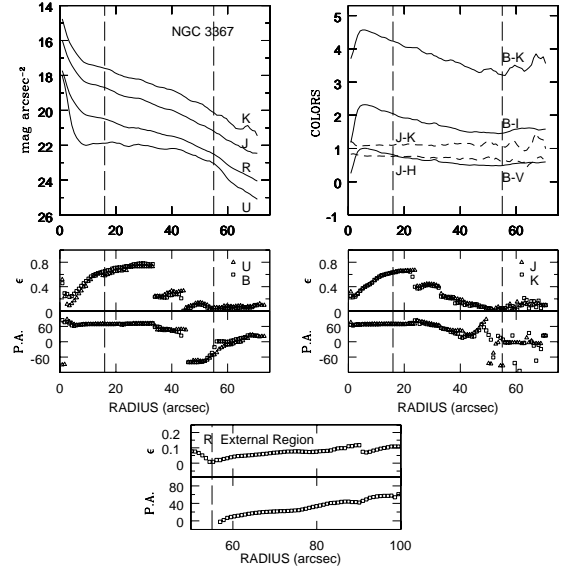}
\caption{Surface Photometry of NGC 3367. Upper panels: the $U$ to $K$ band Surface Brightness profiles and the 
corresponding color profiles. Middle panels: ellipticity and P.A. radial profiles for $U$, $B$, $J$, and $K$ bands. 
Lower panel: external ellipticity and P.A. radial profiles from a deep $R$ band image. The vertical lines in each panel 
indicate the end of the bar and the position of the truncation radii (from the optical images) respectively.}
\label{figura2}
\end{figure}

%\clearpage

\begin{figure}
\plotone{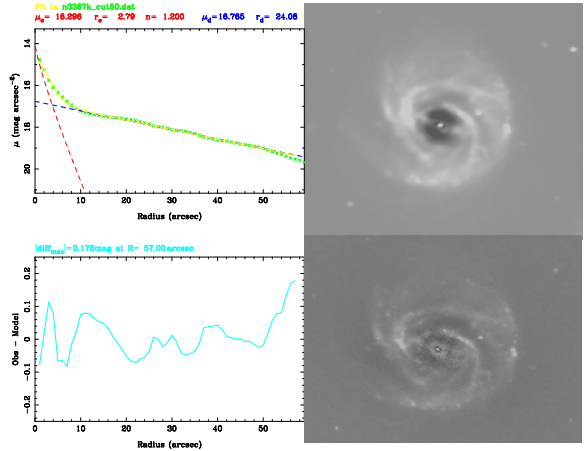}
\caption{Left panel: Best 1-D S\'ersic $+$ Exponential fit to the $K$ band surface brightness 
profile. Right panel: Residual image after subtracting the best 2-D S\'ersic $+$ Exponential fit.
Lower-right panel: Residual image after subtracting the best 2-D S\'ersic $+$ Bar $+$ Exponential fit.}
\label{figura3}
\end{figure}

%\clearpage

\begin{figure}
\plotone{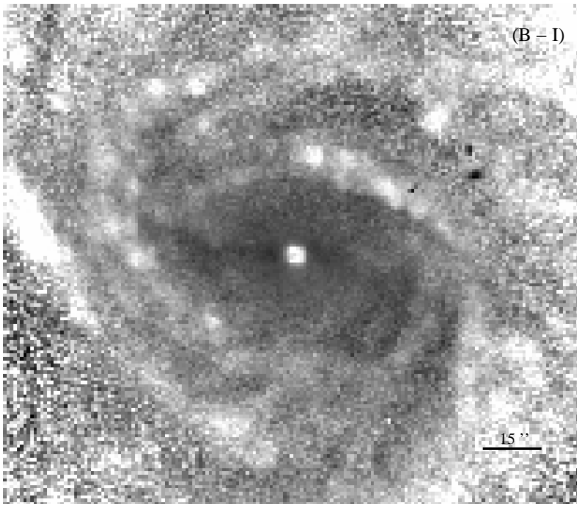}
\caption{A close-up of the inner region of NGC 3367 as seen through a $B-I$ color map. Dark regions represent redder colors. 
The dust lane can be 
appreciated as a dark curved feature running from east to west of the nucleus. It appears slightly offset from the bar major 
axis toward its leading edge. $1''$ = 210 pc. The nucleus and part of the arms are appreciated as gross bluer features.}
\label{figura4}
\end{figure}

%\clearpage

\begin{figure}
\plotone{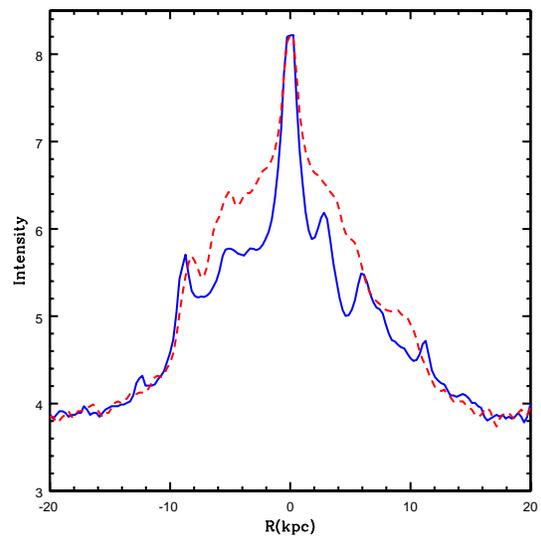}
\caption{$K$ band Intensity profile along and  perpendicular to the bar (dashed and solid lines respectively). The central peak shows 
the unresolved nucleus. Peaks in the profiles located at around 4 kpc show the position of spiral arms which are asymmetric. At 6 kpc 
the profiles reach a ring. Beyond 10 kpc the profiles in orthogonal directions are similar, suggesting that the disk is close 
to axi-symmetric.}
\label{figura5}
\end{figure}

%\clearpage

\begin{figure}
\plotone{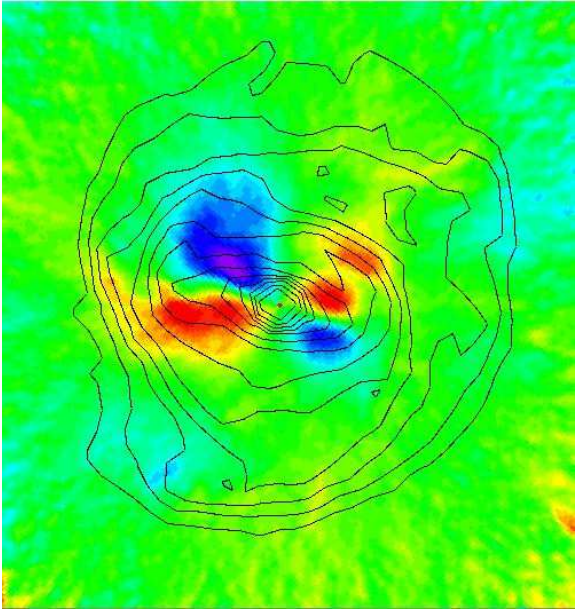}
\caption{The tangential to average radial force ratio map for NGC 3367. The color code illustrates regions, that depending on 
quadrant relative to the bar, can be negative (blue clouds) or positive (red clouds) due to the change of sign of the tangential force. 
The maximum values of the ratio in the red clouds are of the order of +0.42. The maximum values in the upper-left blue cloud are -0.51, while 
those in the lower-right blue cloud are -0.42. Notice the asymmetry in the color distribution along the quadrants. The contours from an 
$R$ band image are overplotted.}
\label{figura6}
\end{figure}

%\clearpage

\begin{figure}
\plotone{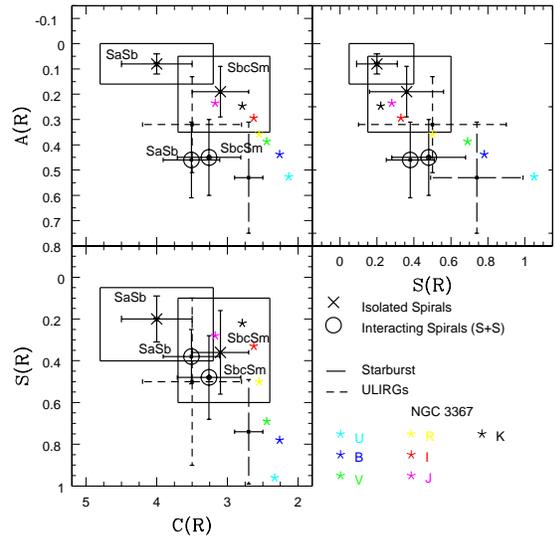}
\caption{$CAS$ Parameter space and the loci of NGC 3367. The $UBVRIJHK$ symbols indicate its wavelength-dependent position. 
Black boxes show the maximum variation of the observed $CAS$ parameters from $B$ to $K$ band for a sample of isolated galaxies 
of Sa-Sb and Sbc-Sm morphological types (Hern\'andez-Toledo et al. 2007; Hern\'andez-Toledo \& Ortega-Esbr\'{\i} 2008). 
Barred circles show the expected loci for interacting (S+S) pairs (Hern\'andez-Toledo et al. 2005). The expected 
loci for Starbursts and ULIR galaxies is taken from C03.}
\label{figura7}
\end{figure}

%\clearpage

\begin{figure}
\plotone{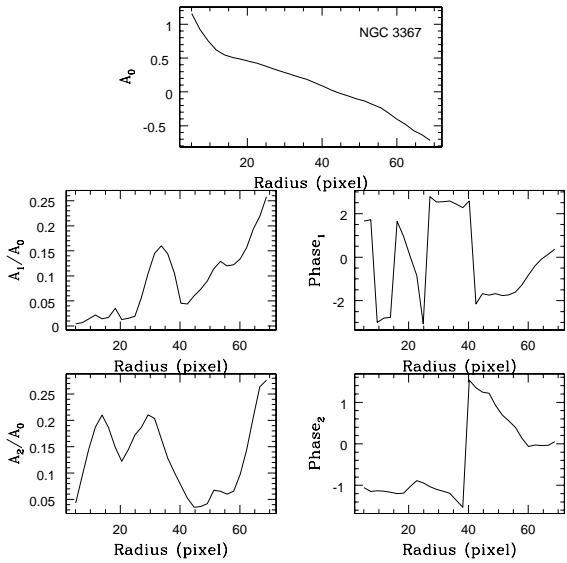}
\caption{Amplitude and phase of Fourier $m=0-1,2$ modes in the $J$ band. Radius is plotted in pixel units. The $J$ band scale-length = 22.54 
pix (19.16 arcsec).}
\label{figura8}
\end{figure}

%\clearpage

\begin{figure}
\plotone{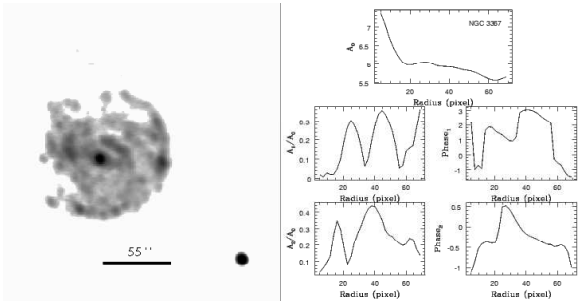}
\caption{Left panel: The moment 0 $H\alpha$ image from the collapsed Fabry-Perot data cube. Right panel: The corresponding $H\alpha$ $m=1,2$ 
Fourier modes.} 
\label{figura9}
\end{figure}

%\clearpage

\begin{figure}
\plotone{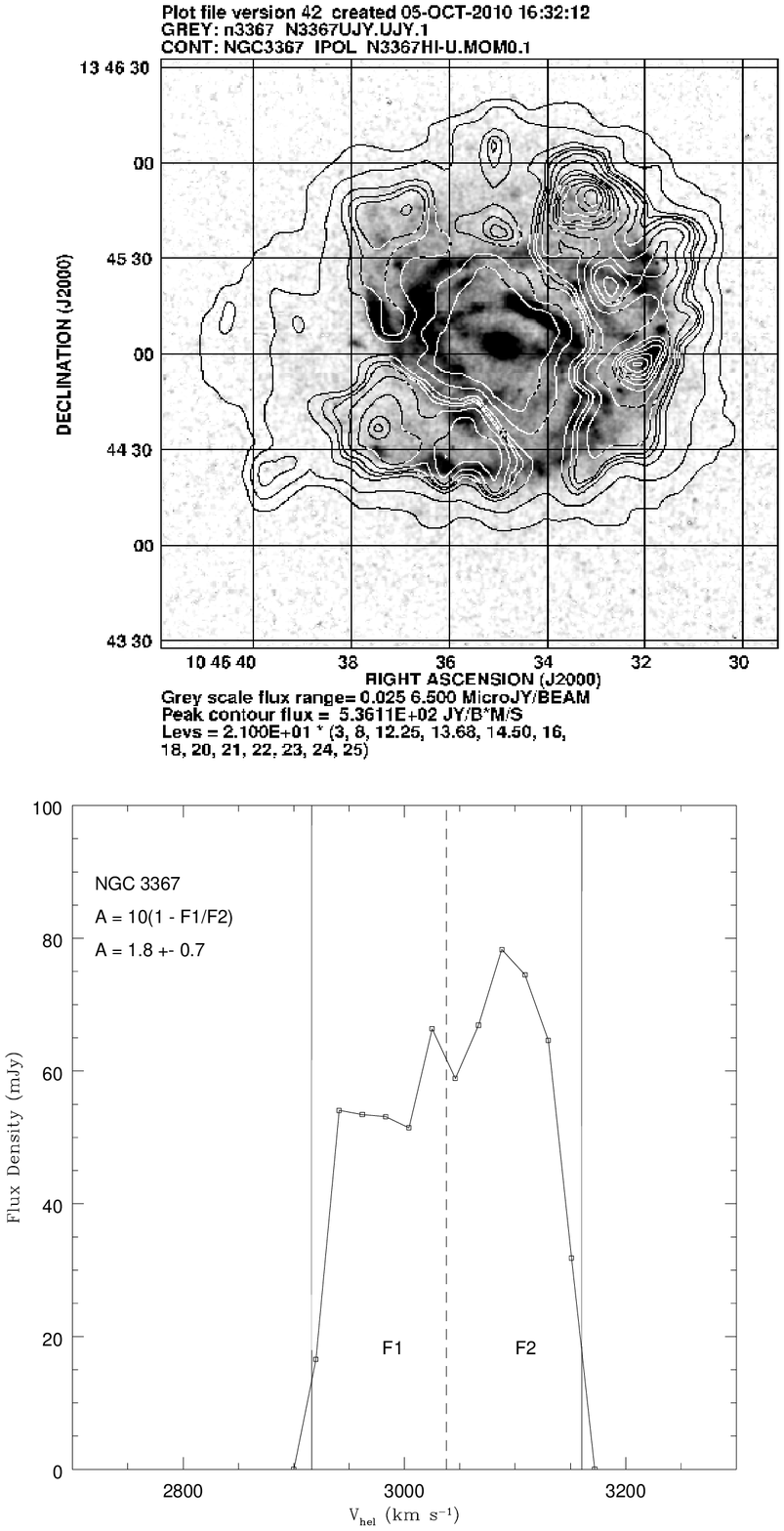}
\caption{Upper panel: The HI intensity map of NGC 3367 from the moment 0 VLA data cube (Bravo-Alfaro et al. in preparation), superposed 
on a $U$ band image. Lower panel: the corresponding integrated HI velocity profile. The center, the extreme values at the 20\% level and 
the integrated flux of each HI horn F1 and F2, are used to estimate the HI asymmetry.}
\label{figura10}
\end{figure}

%\clearpage

\begin{figure}
\plotone{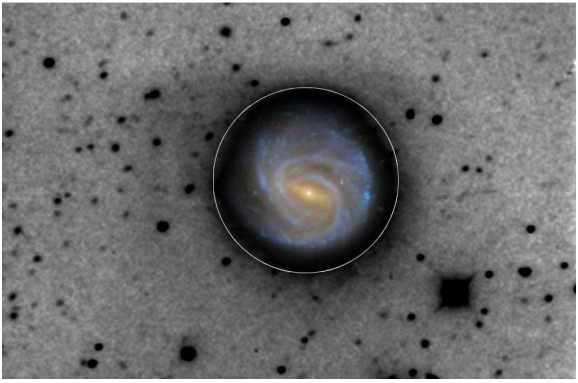}
\caption{
A 1.4 hour $R$ band logarithmic-scaled image of NGC 3367 (in dark). The inset in the central region is the standard
color-coded SDSS image of the galaxy. The circle at $\sim 90$ arcsec illustrates the position of the
region with $S/N = 3$. Beyond that circle the signal seems to be marginal (see discusion in the text).}
\label{figura11}
\end{figure} 

%\clearpage

\begin{figure}
\plotone{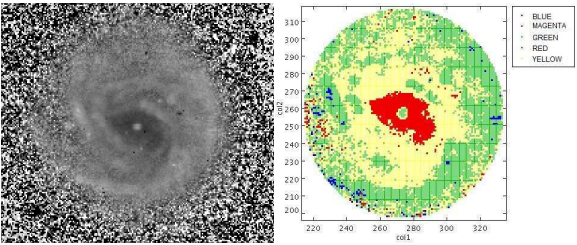}
\caption{($B-V$) color index map of NGC 3367 (upper left panel) and its pixel representation (upper right panel). 
Pixels are codded as follows: blue (0 $<$ ($B-V$) $<$ 0.5), green (0.5 $<$ ($B-V$) $<$ 0.75), yellow (0.75 $<$ ($B-V$) $<$ 1.0), 
red ($B-V$) $>$ 1.0.}
\label{figura12}
\end{figure}

\end{document}